\documentclass[trackchanges,twocolumn]{aastex701}
\usepackage{CJK}
\usepackage{soul}
\usepackage[normalem]{ulem}

\newcommand{\red}[1]{}
\newcommand\HIt{{H\,\small{I}}}
\definecolor{ForestGreen}{RGB}{34,139,34}
\def\tw#1 {{\textcolor{ForestGreen}{#1}}\ }

\begin{document}
\begin{CJK*}{UTF8}{gbsn}
\title{
{
Even central galaxies feel their environment: cosmic siphoning of cool gas accretion}}
\author[0000-0002-8046-984X]{Ke Xu (许可)}
\affiliation{School of Astronomy and Space Science, Nanjing University, Nanjing, Jiangsu 210023, China}
\affiliation{Key Laboratory of Modern Astronomy and Astrophysics, Nanjing University, Ministry of Education, Nanjing 210023, China}
\email{kexu@nju.edu.cn}
\author[0000-0002-2504-2421,sname=Wang,gname=Tao]{Tao Wang (王涛)}
\affiliation{School of Astronomy and Space Science, Nanjing University, Nanjing, Jiangsu 210023, China}
\affiliation{Key Laboratory of Modern Astronomy and Astrophysics, Nanjing University, Ministry of Education, Nanjing 210023, China}
\email[show]{taowang@nju.edu.cn}  
\begin{abstract}
{Identifying the key processes regulating cool gas accretion in galaxies is essential for understanding the baryon cycle within galactic ecosystems. Recent studies indicate that the most fundamental internal process is likely the competition between halos and central black holes. This is revealed through a universal relation between increasing black hole masses and decreasing cool gas fraction ($\mu_{\rm HI} \equiv M_{\rm HI}/M_{\star} \propto M_{\rm BH}^{-0.6}$).
Here we further explore the primary external effects on cool gas accretion in central galaxies. We find that galaxies deviating significantly from the $\mu_{\rm HI}-M_{\rm BH}$ relation are overwhelmingly isolated central galaxies without substantial satellites. These isolated, gas-poor centrals exhibit systematically higher black-hole-to-stellar-mass ratios and more prominent bulges. At larger scales, they are preferentially located near massive neighboring halos.
We suggest that this may reflect a ``cosmic siphoning" scenario, \red{in which the cool-gas supply of isolated centrals }{whereby the cool-gas supply in a subset of isolated centrals is curtailed by nearby massive halos. In contrast, group centrals are continually} replenished by their satellites and by gas flowing along the cosmic web. \red{This picture offers a plausible explanation both for the different gas fractions of isolated and group centrals and for the gas deficiency of the isolated galaxies that sit close to a massive halo, complementing the internal regulation by supermassive black holes and embodies a cosmic ``rich-get-richer" paradigm.}
{This model naturally explains the prevalence of gas-poor isolated central galaxies near massive halos. It complements internal regulation by supermassive black holes and suggests a cosmic 'rich get richer' scenario.}
}


\end{abstract}


\section{Introduction} \label{sec:intro}
Composed of stars, a multi-phase medium, dark matter and often, a supermassive black hole (SMBH), galaxies evolve as ecosystems regulated by various physical processes operating across different spatial and temporal scales~\citep{Tumlinson+2017,Naab+2017}. A central goal of extragalactic astronomy is to identify the most fundamental mechanisms governing this evolution. In particular, because cool gas serves as the primary reservoir for star formation and a key component of the baryon cycle,  what fundamentally regulates the cool gas content in galaxies is a key question in galaxy formation and evolution.

As in other complex systems, gas recycling in galactic ecosystems is driven by the interplay of internal and external processes. In terms of the main internal mechanisms, dark matter halos are expected to play an important role by setting up gravitational potential, which determines the  rates and modes of baryonic accretion~\citep{Kerevs+2005,Dekel+2006,Dekel+2009,Guo+2021,LiXiao+2022,DouJing+2025}. 
However, it has long been shown that, without introducing additional feedback mechanisms, gas in dark matter halos would cool too efficiently and form stars in too great a quantity, raising the so-called ``overcooling problem"~\citep{Benson+2010,Somerville+2008}. 
Feedback is essential to solve this overcooling problem by providing a source of heating or energy injection to disrupt the cooling flow and/or eject gas, and suppress star formation.
At least for massive galaxies, this source of feedback is believed to be Active Galactic Nuclei (AGN)~\citep{Somerville+2008,McCarthy+2011,Fabian+2012}. Previous observational evidence for AGN feedback affecting the gas recycling in galaxies mostly exists for central galaxies in clusters, based on the discovery of vast cavities associated with radio jets in the hot gas of galaxy clusters~\citep{Rafferty+2006,McNamara+2007,Fabian+2012}.
More recently, \citet{WangTao+2024} demonstrated that, for general group central galaxies and among a wide range of galaxy properties, the cool gas fraction (as approximated by $\mu_{\rm HI} \equiv M_{\rm HI}/M_{\star}$) correlates most tightly with the mass of central supermassive black holes (SMBHs). They further interpreted this relation as arising from the balance between the halo binding energy and the cumulative energy released by the SMBH (as well as the energy conversion efficiencies for SMBHs). This provides direct evidence on gas recycling in galaxies as a balanced, self-regulating cycle between cosmic accretion (and cooling) and heating feedback, driven by dark matter halos and SMBHs.


In terms of external mechanisms, satellite galaxies are well known to suffer from group- and cluster-related environmental processes, including tidal interactions, harassment, and ram-pressure stripping up to high redshift~\citep{Gunn+1972,Toomre+1972,Moore+1998,Larson+1980,LiCheng+2012,WangJing+2020,Alberts+2022,Boselli+2022,Boselli+2019,Noble+2019,Cramer+2023}. 
Central galaxies, by contrast, are exempt from the violent intra-group processes that affect satellites. Yet they are not immune to their surroundings: their location within the cosmic web (e.g., filaments, tendrils, voids) determines the accessibility of external gas reservoirs on Mpc scales. Meanwhile, the host halo mass sets the characteristic dark-matter accretion rate~\citep{McBride+2009}, which in turn governs the gravitational potential that draws in baryonic gas. The interplay between these environmental and halo-mass channels regulates the net gas accretion onto central galaxies. For example, \cite{Janowiecki+2017} have found that isolated central galaxies (field galaxies) \red{are more gas-poor}{have an average lower atomic gas fraction} than group central galaxies, especially at lower stellar mass. After matching star-formation rates, \cite{YanShulan+2026} still consistently found an excess of the gas fraction in group central galaxies compared to isolated galaxies.
The gas stripping from the large-scale structures could account for this discrepancy, evidenced by increasing \HIt{}-deficiency toward filament spines in the field~\citep{CroneOdekon+2018,Hoosain+2024}.

Most previous \red{studies addressed this gas deficiency}{work studied the atomic gas content} by examining the relation among the cool gas fraction, stellar masses, and galaxy optical colors (or specific star formation rates) ~\citep{ZhangWei+2009,ZuYing+2020,Catinella+2010,WangJing+2017,Catinella+2018,LiXiao+2022}. Here building upon the previous results from ~\cite{WangTao+2024}, which suggest that the balance between halos and SMBHs is the most fundamental process regulating the cool gas fraction in group central galaxies, we explore whether the differences of $\mu_{\rm HI}$ between isolated and group central galaxies~\citep{Janowiecki+2017,YanShulan+2026} could be explained by this $\mu_{\rm HI}-M_{\rm BH}$ relation or \red{whether} isolated galaxies are \red{intrinsically more gas-poor than}{still different from} group central galaxies at fixed $M_{\rm BH}$. 
If the latter is the case, while central galaxies are shaped primarily by internal SMBH feedback, large-scale environment (beyond group scales) still plays a non-negligible secondary role.
Moreover, previous work has mainly disentangled the effects of large-scale structure by identifying filamentary structures from the galaxy distribution. Here, we directly link environmental effects to halo properties, e.g., whether the influence of large-scale structure arises from the competitive accretion with nearby halos.

This paper is organized as follows: Sec.~\ref{sec2} introduces the data used in this paper and the derivation of the black hole masses for our sample galaxies. Sec.~\ref{sec3} shows the relation between the atomic gas and black hole masses for isolated and group central galaxies, and the internal and external dependence of the gas deficiency\red{ in isolated galaxies}. Sec.~\ref{sec4} discusses the regulation of cool gas in different environments. And we conclude with Sec.~\ref{sec5}. 

Throughout this paper, we use $\rm\Lambda$CDM cosmology with $\Omega_{\rm M}=0.3$, $\Omega_{\rm \Lambda}=0.7$ and $H_0=70~\rm km/s$.

\section{Data}\label{sec2}
\subsection{Galaxy sample and atomic gas information}

In this study, we use the galaxy sample from MaNGA survey \citep{Bundy+2015}, which has a flat distribution in the stellar mass along with abundant auxiliary data and analyses, especially the atomic gas observation from \HIt{}-MaNGA survey~\citep{Masters+2019,Stark+2021}. The \HIt{}-MaNGA survey observes the galaxies with Arecibo \citep{Haynes+2011,Haynes+2018} and Green Bank Telescope (GBT), which lack the spatial resolution and could have confusion from projected nearby galaxies. We only use the galaxies with confusion probability $(\text{\bf conf\_prob})$ $<0.5$ reported by \cite{Stark+2021}. 
And we recalculate the \HIt{} upper limits for \HIt{}-undetected galaxies as in \cite{WangTao+2024} via the relation between the line width and the stellar mass rather than adopting the default 200~km/s line width.
To further study the evolution of galaxies, we attempt to eliminate the satellites which suffer from group/cluster effects, by only selecting the group central galaxies (GCs) and isolated central galaxies (ICs) from the group catalog~\citep{Argudo+2015,Yang+2007}. ICs and GCs are defined as the galaxies without a satellite and the brightest galaxy in the group with satellite galaxies, respectively.

We also examine our results with the galaxy sample from xGASS survey~\citep{Catinella+2010,Catinella+2018} applying the same selection criteria. The main results are consistent and are shown in the Appendix~\ref{appendix}.

\subsection{Basic galaxy properties}
The redshift and distance information of MaNGA galaxies are from the summary file of MaNGA Data Analysis Pipeline \citep[DAP;][]{Westfall+2019}. We exclude galaxies with $z>0.035$ to avoid a bias to gas-rich galaxies at longer distance. We collect the stellar masses and star-formation rates (SFRs) from GSWLC-2 catalog \citep{Salim+2016,Salim+2018}, which includes the mid-IR photometry in the SED fitting and would provide more precise estimations. The velocity dispersion measurement within galaxy effective radius ($\sigma_\star$) and axis ratio ($b/a$) are from \cite{Zhu+2023}, the morphological T-Type is from~\cite{Dominguez+2018}, and the light-weighted stellar ages within the effective radius ($\rm Age_{r_e,LW}$) are from the \textsc{FIREFLY} pipeline~\citep{Neumann+2022}.

\setcounter{footnote}{0}
\subsection{The black hole masses}
\begin{figure}[t]
\centering
\includegraphics[width=0.49\textwidth]{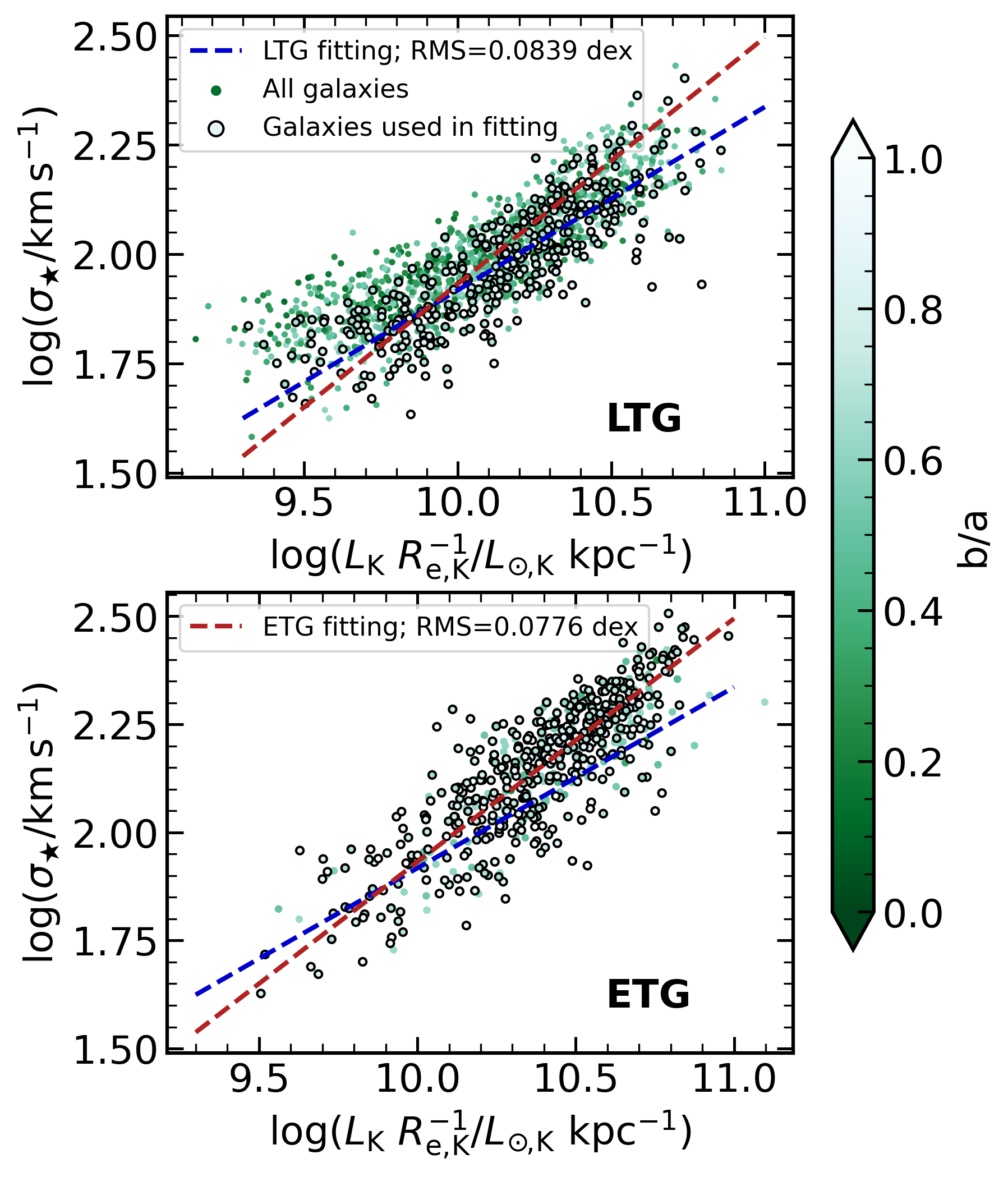}
\caption{\textbf{The calibrated relation among the stellar velocity dispersion, the K-band luminosity and the K-band effective radius.} The data points are color-coded by their axis ratios. The dashed blue and red lines are the best-fitted relation for LTGs and ETGs, respectively. The best-fitted relations are $\log\sigma_\star=0.42(0.02)\log(L_{\rm K}/R_{\rm e,K})-2.27(0.19)$ (LTG) and $\log\sigma_\star=0.56(0.02)\log(L_{\rm K}/R_{\rm e,K})-3.70(0.22)$ (ETG).}\label{fig1}
\end{figure}

While the direct black hole masses can only be derived in a small portion of nearby galaxies, some empirical relations are proposed, e.g. the relation between the supermassive black hole mass and stellar velocity dispersion ($M_{\rm BH}\,-\,\sigma_\star$)~\citep{Ferrarese+2000,Gebhardt+2000,Tremaine+2002,McConnell+2013,Kormendy+2013}. Including other galaxy properties in this relation could not produce a tighter correlation \citep{Beifiori+2012}. For late-type galaxies (LTGs), the rotation velocity would contribute to the $\sigma_\star$, which leads to the negative offsets from $M_{\rm BH}\,-\,\sigma_\star$ relation in high-inclination cases~\citep{Xiao+2011}. Therefore we need to better derive the velocity dispersion in high-inclination galaxies. The elliptical galaxies obey tight Fundamental Plane (FP) relation~\citep{Djorgovski+1987,Dressler+1987}. If we use the near-infrared photometry, spiral galaxies would have minimized variation of mass-to-light ratio~\citep{Falcon-Barroso+2011,Meidt+2014,Norris+2014} and could be incorporated into the FP relation~\citep{Bosch+2016}. Moreover, \cite{ZhuKai+2024} have also presented the FP relation for MaNGA late- and early-type galaxies (ETGs) respectively with the r-band luminosity. Here we refit the FP relation among the K-band luminosity ($L_{\rm K}$), effective radius ($R_{\rm e,K}$) from 2MASS~\citep{Skrutskie+2006,Jarrett+2000,2MASS}\footnote{\url{https://catcopy.ipac.caltech.edu/dois/doi.php?id=10.26131/IRSA97}}, and the velocity dispersion in the effective radius ($\sigma_{\rm \star}$) from~\cite{Zhu+2023} only for galaxies with axis ratio $b/a>0.64$ ($\sim$ 50\,$\deg$ inclination assuming the intrinsic disk thickness of 0.2). ETGs and LTGs are categorized with a division of T-Type~$=0$. We apply this relation to the whole galaxy sample and finally calculate the $M_{\rm BH}$ with equation 2 in \cite{Bosch+2016} as $\log(M_{\rm BH}/{M_{\odot}})=8.32+5.35\,\log(\sigma_{\rm \star}/200~{\rm km~s^{-1}})$. The best-fitted result is shown in Fig.~\ref{fig1}.

Our final sample incorporates the galaxies with valid $M_{\rm BH}$, group information and basic galaxy properties, which returns 1060 ICs (577 \HIt{}-detected and 483 \HIt{}-undetected) and 393 GCs (214 \HIt{}-detected and 179 \HIt{}-undetected).

\section{Results}\label{sec3}

\begin{figure*}[t]
\centering
\begin{minipage}[t]{0.99\textwidth}
\includegraphics[width=0.99\textwidth]{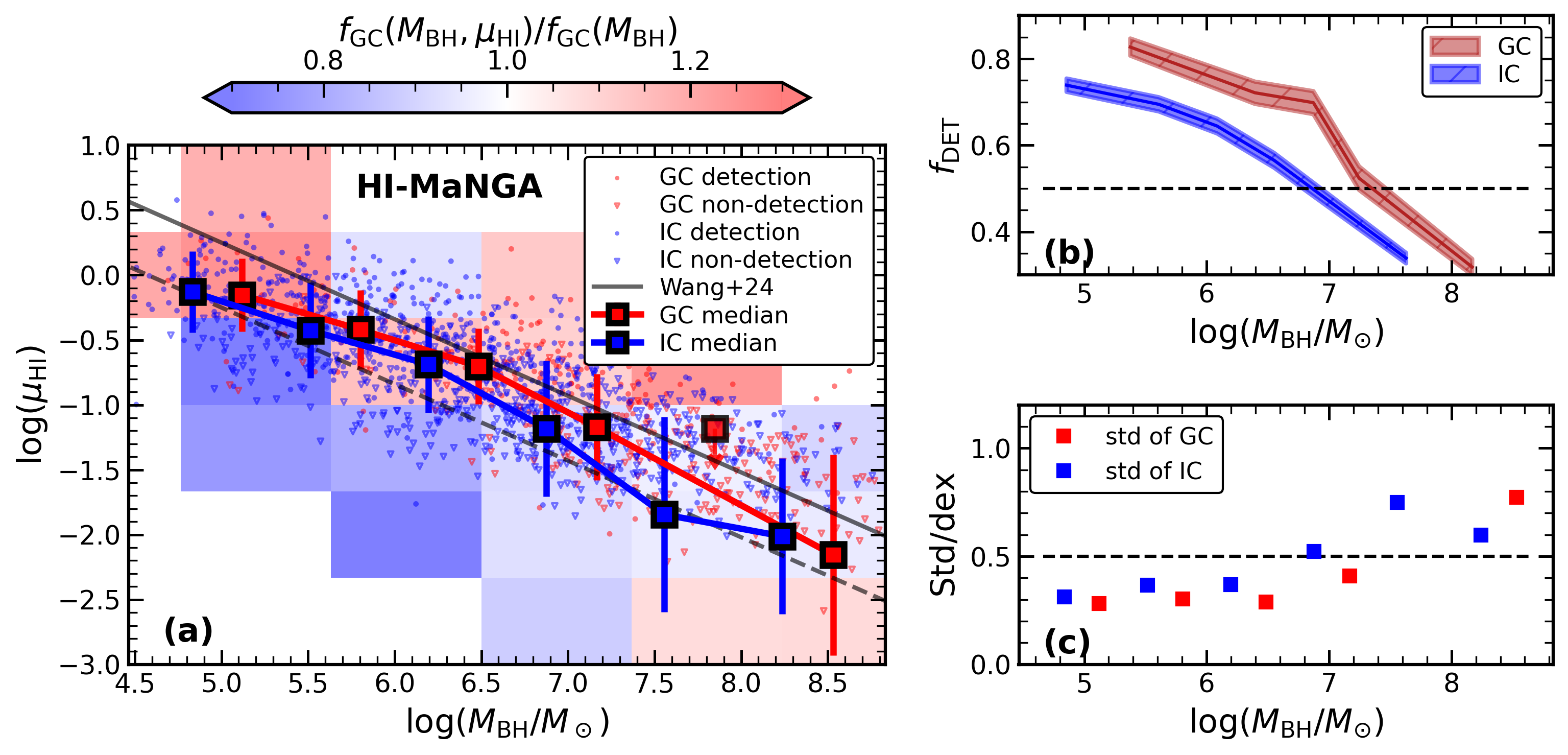}
\end{minipage}
\begin{minipage}[t]{0.99\textwidth}
\includegraphics[width=0.99\textwidth]{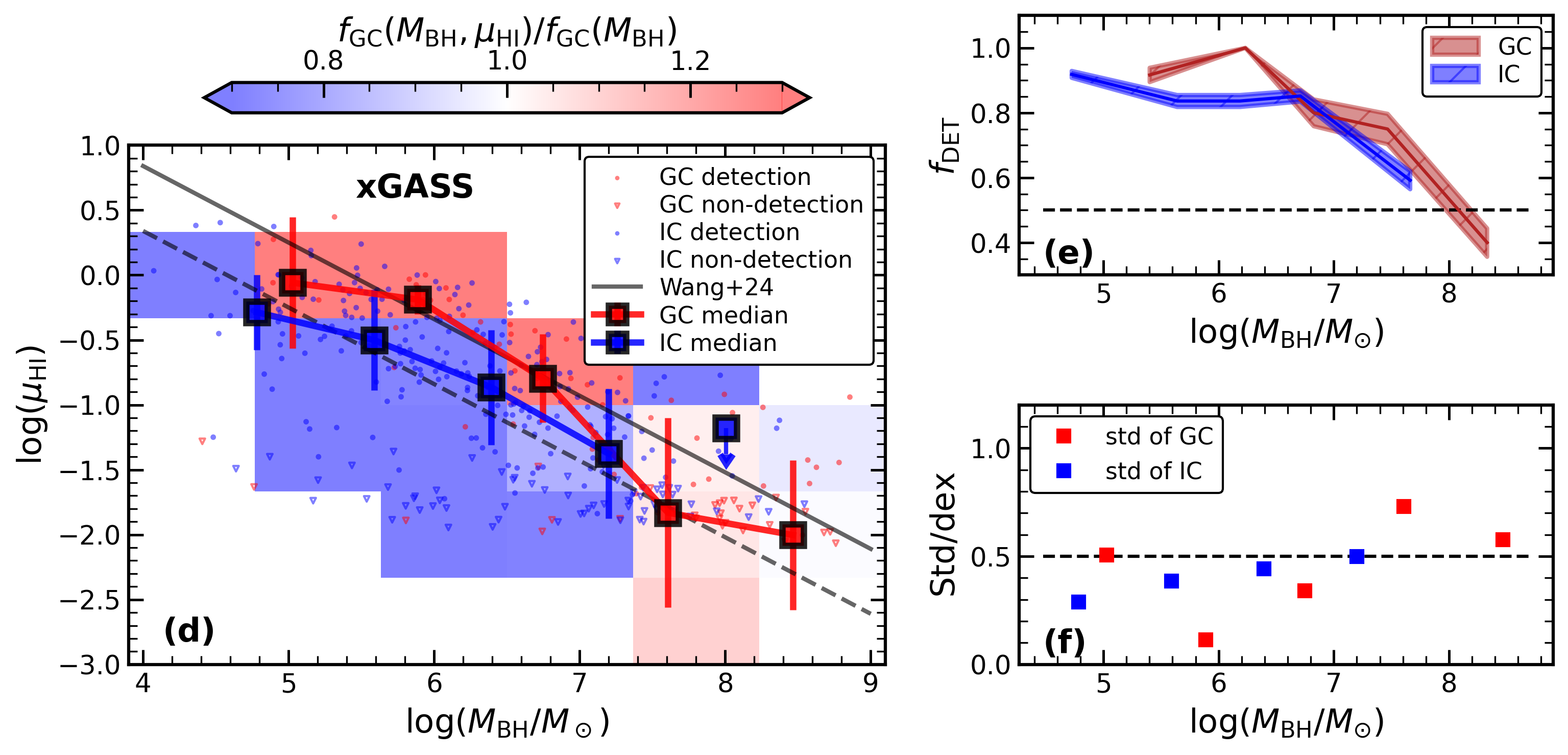}
\end{minipage}
\caption{\textbf{The relation between $\mu_{\rm HI}$ and $M_{\rm BH}$ for group centrals (GCs) and isolated centrals (ICs) in H\textsc{I}-MaNGA (upper a-c) and xGASS (lower d-f).} In the top three panels: (a) is the scatter plot between $\mu_{\rm HI}$ and $M_{\rm BH}$. The background color map represents the ratio between the fraction of GCs ($f_{\rm GC}$) at given $\mu_{\rm HI}$ \& $M_{\rm BH}$, and the $f_{\rm GC}$ at given $M_{\rm BH}$ (only grids with $\geq5$ counts are shown with color). The black solid line is the relation from \cite{WangTao+2024} and corresponding dashed black line shifts this relation 0.5~dex below for reference. (b) shows the \HIt{}-detection fraction of GCs and ICs. (c) shows the standard deviation of the median in each $M_{\rm BH}$ bin corresponding to the errorbar of the median in (a). (d)-(f) are the same as (a)-(c) except for that the sample is changed to xGASS galaxies. }\label{fig2}
\end{figure*}

\begin{figure*}[t]
\centering
\includegraphics[width=0.99\textwidth]{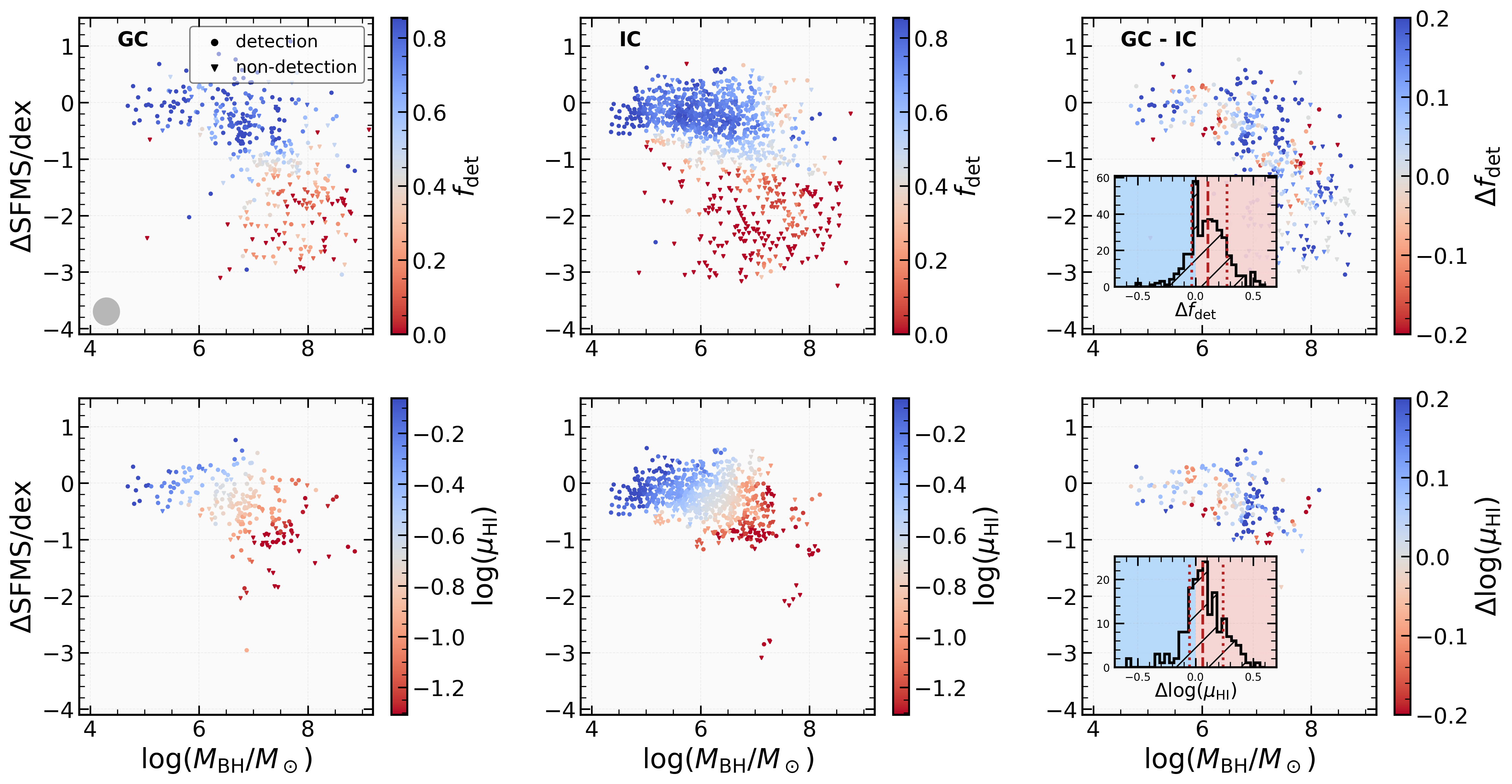}
\caption{\textbf{From the left to the right, we show the HI-detection fraction (upper) and the gas fraction (lower) on the $\Delta {\rm SFMS}-M_{\rm BH}$ plane for GCs, ICs and their corresponding differences.} The first two columns show the $f_{\rm det}$ (upper row) and $\mu_{\rm HI}$ (lower row) as the color coding for GCs and ICs. The ellipse in the lower-left of the first panel shows the aperture size used to smooth the color coding. The third column shows the difference between GCs and ICs (controlling $M_{\rm BH}$ and $\Delta \rm SFMS$). For each GC, we identify isolated counterparts within a 0.25~dex vicinity on the $\Delta\rm SFMS$-$M_{\rm BH}$ plane and calculate the difference to the median value of the matched counterparts. The inserted panel in the third column further shows the 1-D distribution of this color difference derived in the third columns and the red dashed and dotted lines denote the 50th, 16th and 84th percentiles of the distribution.}\label{fig3}
\end{figure*}

\subsection{\red{Gas deficiency in ICs compared to GCs}{Differences in \HIt{} content between ICs and GCs}}
We extend the $\mu_{\rm HI}$ - $M_{\rm BH}$ relation in \cite{WangTao+2024} to ICs in this work.
The extended sample (for both \HIt{}-MaNGA and xGASS) is shown in Fig.~\ref{fig2}. The median values are given by Kaplan-Meier estimator taking into account the \HIt{}-undetected galaxies. For those bins with insufficient detected galaxies (low \HIt{}-detection fraction), we use the 84th percentile from the survival function as the upper limit of the median value.

For \HIt{}-MaNGA sample, ICs exhibit slightly lower gas fraction than GCs while the scatter is quite large given the relatively low observational sensitivity. The lower gas content in ICs is more clearly observed with xGASS survey \citep[in panel d of Fig.~\ref{fig2}, and also][]{Janowiecki+2017}, reaching a deeper detection limit of $\mu_{\rm HI}\sim2-5\%$. In the panel b (and also e) of Fig.~\ref{fig2}, ICs have a lower \HIt{}-detection ratio than GCs, which indicates that more ICs have gas fraction lower than the detection limit and exhibit larger gas deficiency. 
We note that in \HIt{}-MaNGA sample, GCs are also slightly lower than the relation in ~\cite{WangTao+2024}, because the deduction of rotation velocity in the velocity dispersion introduces smaller $M_{\rm BH}$, which is more obvious for the low-$M_{\rm BH}$ LTGs (Fig.~\ref{fig1}). The other reason of this deviation from \cite{WangTao+2024} is the different sources of stellar masses: \cite{WangTao+2024} matched from MPA-JHU catalog~\citep{Kauffmann+2003,Brinchmann+2004,Tremonti:2004} and this work utilized from \cite{Salim+2018}, which have systematic offset.

For comparison, we also show the $\mu_{\rm HI}-M_\star$ relation in the Appendix (Fig.~\ref{figA1}). To compare the $\mu_{\rm HI}-M_\star$ and $\mu_{\rm HI}-M_{\rm BH}$ relation, we divide the sample into bins equally and use the Kaplan-Meier estimator to derive the standard deviation in each bin, shown in the panel c and f in Figs.~\ref{fig2}~\&~\ref{figA1}. The $\mu_{\rm HI}-M_\star$ relation exhibits larger scatter than the $\mu_{\rm HI} - M_{\rm BH}$ relation, as found in \cite{WangTao+2024}. The main reasons for the differences of the two relations are the mixed galaxy morphology at given $M_\star$~\citep{WangTao+2024} and the relatively narrow dynamic range of $M_\star$. Given the fact that $M_{\rm BH}$ has tighter correlation with the gas fraction, as the dominant property in the galaxy evolution~\citep{Bluck+2020,WangTao+2024}, in the following we focus on the analysis from the $M_{\rm BH}$ perspective. We note, however, that $M_{\rm BH}$ in this work is inferred from galaxy luminosities and sizes, calibrated against velocity dispersion. The effects attributed to black holes may therefore be intertwined with other galactic properties (e.g., stellar surface density). Nevertheless, the following main results remain unchanged when examined from the perspective of $M_\star$.

Defining galaxies that are more than 0.5~dex below the relation in \cite{WangTao+2024} as outliers, we observe that 84\% of these outliers are isolated (89\% for xGASS sample). This is consistent with the recent finding of \cite{Deo+2026} that recently quenched ellipticals are isolated. Besides, the outliers in ICs mostly have $\log(M_{\rm BH}/{M_\odot})<7.0$, indicating that they could experience strong gas removal events at low $M_{\rm BH}$. 

The large isolated occupation fraction in outliers could be partly attributed to the larger sample size of ICs. 
To further compare the outlier fraction, we calculate the ratio between the fraction of GCs at given $\mu_{\rm HI}$ \& $M_{\rm BH}$, and the fraction of GCs at given $M_{\rm BH}$ (shown in the background color map in panel a~\&~c of Fig.~\ref{fig2}). For \HIt{}-undetected galaxies, we assume they have a uniform probability distribution in the linear space below the upper limits, and account for their fractional contributions to the number counts in each bin.
The group central occupation is lower in the regions of outliers (bluer in the color). \red{This further implies that the field environment of ICs would inherit a lack of gas accretion process or efficient gas removal events, resulting in larger gas deficiency.}{This further implies that the gas deficiency, related to a lack of gas accretion processes or efficient gas removal events, is more frequent in the field environments.}

The star-formation rate (SFR) is not controlled during the above analysis, when some previous work suggested that the color or the SFR is tightly correlated with the gas content~\citep[e.g., ][]{Catinella+2018,Janowiecki+2017}. 
In the following analyses, we show the results using \HIt{}-MaNGA sample in the main text because of larger sample size. And the results using xGASS sample are shown in the Appendix~\ref{appendix}.

To exclude the influence of different  star-formation states, we plot the $\mu_{\rm HI}$ and \HIt{}-detection fraction ($f_{\rm det}$) over the $\Delta {\rm SFMS}-M_{\rm BH}$ plane in Fig.~\ref{fig3}, where $\Delta {\rm SFMS}=\log({\rm SFR/SFR_{MS}}(M_\star))$ and the star-formation rate of the main-sequence galaxy ($\rm SFR_{MS}$) is given by \cite{Renzini+2015}. The color represents the median value within a neighborhood of radius 0.25\,dex, and we present the $\mu_{\rm HI}$ values exclusively for valid estimation from the Kaplan-Meier estimator. The third column in Fig.~\ref{fig3} shows the difference of $\mu_{\rm HI}$ and $f_{\rm det}$ between GCs and ICs, with the inserted panel showing the 1D distribution of the differences from the third column. 
Even after controlling for both $M_{\rm BH}$ and $\Delta\rm SFMS$, ICs consistently show lower \HIt{} detection fraction than GCs.
When altering the color coding to $\mu_{\rm HI}$, despite the median value being reliably obtainable only for main-sequence galaxies and the limited depth of the \HIt{}-MaNGA observation, the result remains consistent: GCs inherently possess a slightly higher {median} gas fraction compared to ICs~\citep[as in][]{YanShulan+2026}. The group environment could be important in the host galaxy evolution on account of the larger binding energy in a larger halo to have sustainable gas reservoir from the multiple-phase gas accretion~\citep[e.g., ][]{Temi+2018,Oppenheimer+2021,Lagos+2022,Fabian+2025}. In recent observations, there is also evidence for rejuvenation of radio activity in GCs~\citep{WangYijun+2025}, especially triggered by \HIt{}-gas inflow ~\citep{Cotton+2020,Mtshweni+2025}.

Throughout this paper we \red{describe}{interpret} the IC-GC offset as a \red{gas deficiency}{lower median gas fraction} in ICs. Because ICs are the dominant galaxy population both in our sample and in the local central galaxies, the same correlation can be also expressed as \red{an excess of cold gas}{a higher median gas fraction} in GCs. In this work, we use ``normal" to denote those galaxies sitting on \cite{WangTao+2024}'s $\mu_{\rm HI}-M_{\rm BH}$ relation calibrated with GC sample. We adopt GCs as the reference population for two reasons. {(i)} The $\mu_{\rm HI}-M_{\rm BH}$ relation is interpreted as the balance between halo binding energy and energy released by the SMBH. At a given stellar mass, ICs are expected to reside in smaller halos, which limits their ability to accrete gas, together with the absence of interactions with satellites. Hence the lower {median} gas fraction would be attributed at least partly to the inaccessibility of gas replenishment rather than the self-regulated results {in a sub population of ICs}. (ii) \red{From the xGASS sample (which has a higher sensitivity) in Figure~\ref{fig2}, the gas-deficit ICs show distinct populations (more obvious at $log(M_{\rm BH}/M_\odot)<6.0$) rather than a continuous distribution below the $\mu_{\rm gas}-M_{\rm BH}$ relation. }
{In the more sensitive xGASS sample shown in Figure~\ref{fig2}, the gas-deficit ICs form a distinct population (particularly at $log(M_{\rm BH}/M_\odot)<6.0$ where $\sim20\%$ ICs have $\log \mu_{\rm HI}<-1$) rather than a continuous distribution below the $\mu_{\rm gas}-M_{\rm BH}$ relation.}
This means that the ICs are heterogeneous internally {and this gas-deficit population is the main contributor to the lower median $\mu_{\rm HI}$ of ICs relative to GCs.} On the contrary, GCs have low outlier fraction which are more homogeneous. And we would further discuss the \red{reasons for gas deficiency in ICs}{cause of the \HIt{}-deficit ICs} in the following sections.

\begin{figure*}[htbp]
\centering
\includegraphics[width=0.99\textwidth]{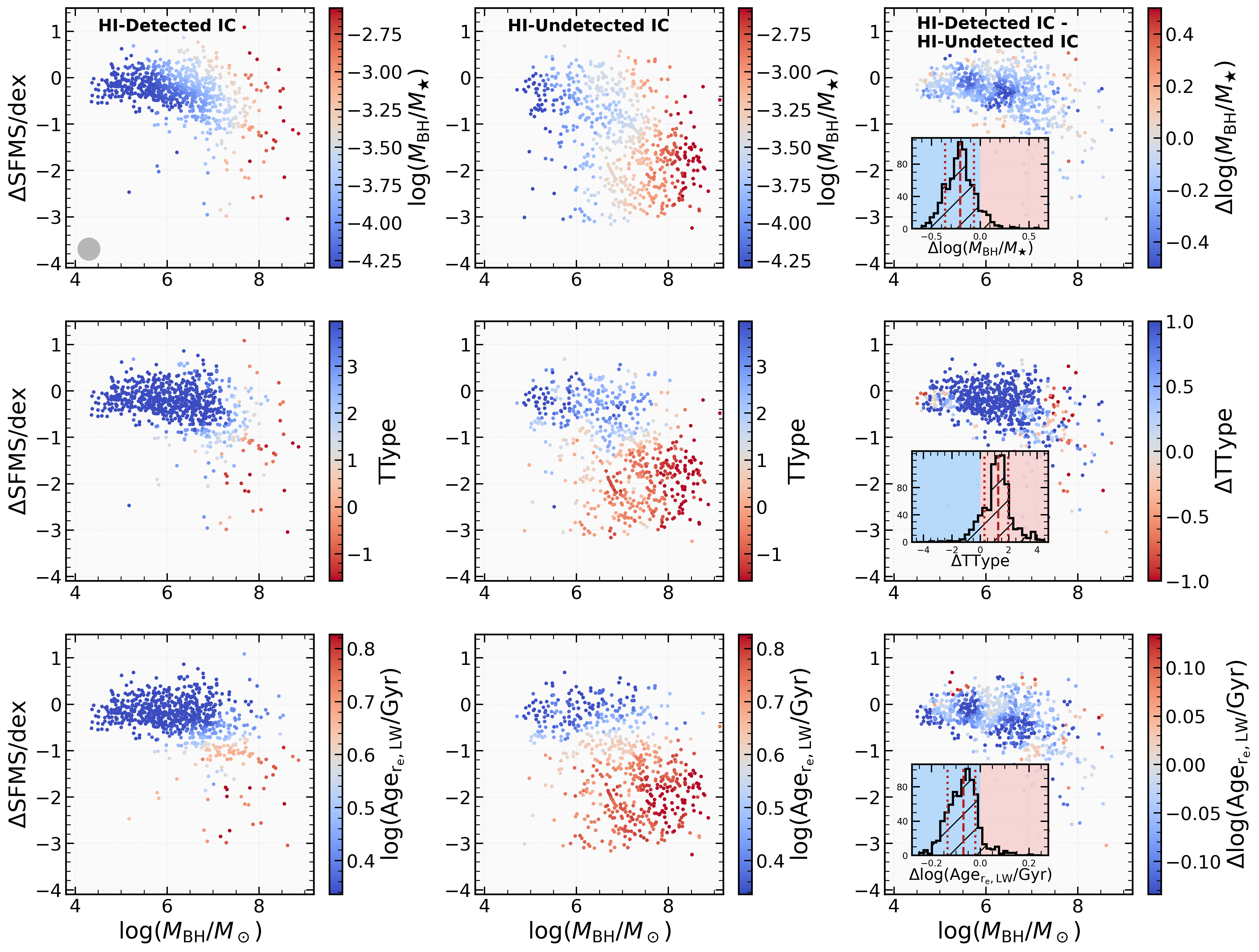}
\caption{\textbf{From the left to the right, we show the $M_{\rm BH}/M_\star$ (the first row), the T-Type (the second row), and the light-weighted stellar ages (the third row) on the $\Delta {\rm SFMS}-M_{\rm BH}$ plane for the \HIt{}-detected ICs, \HIt{}-undetected ICs and their corresponding differences.} The first two columns show the galaxy properties as the color coding for the \HIt{}-detected and \HIt{}-undetected ICs. The ellipse in the lower-left of the first panel shows the aperture size used to smooth the color coding. The third column shows the difference between the \HIt{}-detected and \HIt{}-undetected ICs controlling $M_{\rm BH}$ and $\Delta \rm SFMS$. For each \HIt{}-detected galaxy, we identify \HIt{}-undetected counterparts within a 0.25~dex vicinity on the $\Delta\rm SFMS$-$M_{\rm BH}$ plane and calculate the difference to the median value of the matched counterparts. The inserted panel in the third column further shows the 1-D distribution of this color difference derived in the third columns and the red dashed and dotted lines denote the 50th, 16th and 84th percentiles of the distribution.}\label{fig4}
\end{figure*}

\begin{figure*}[htbp]
\centering
\includegraphics[width=0.99\textwidth]{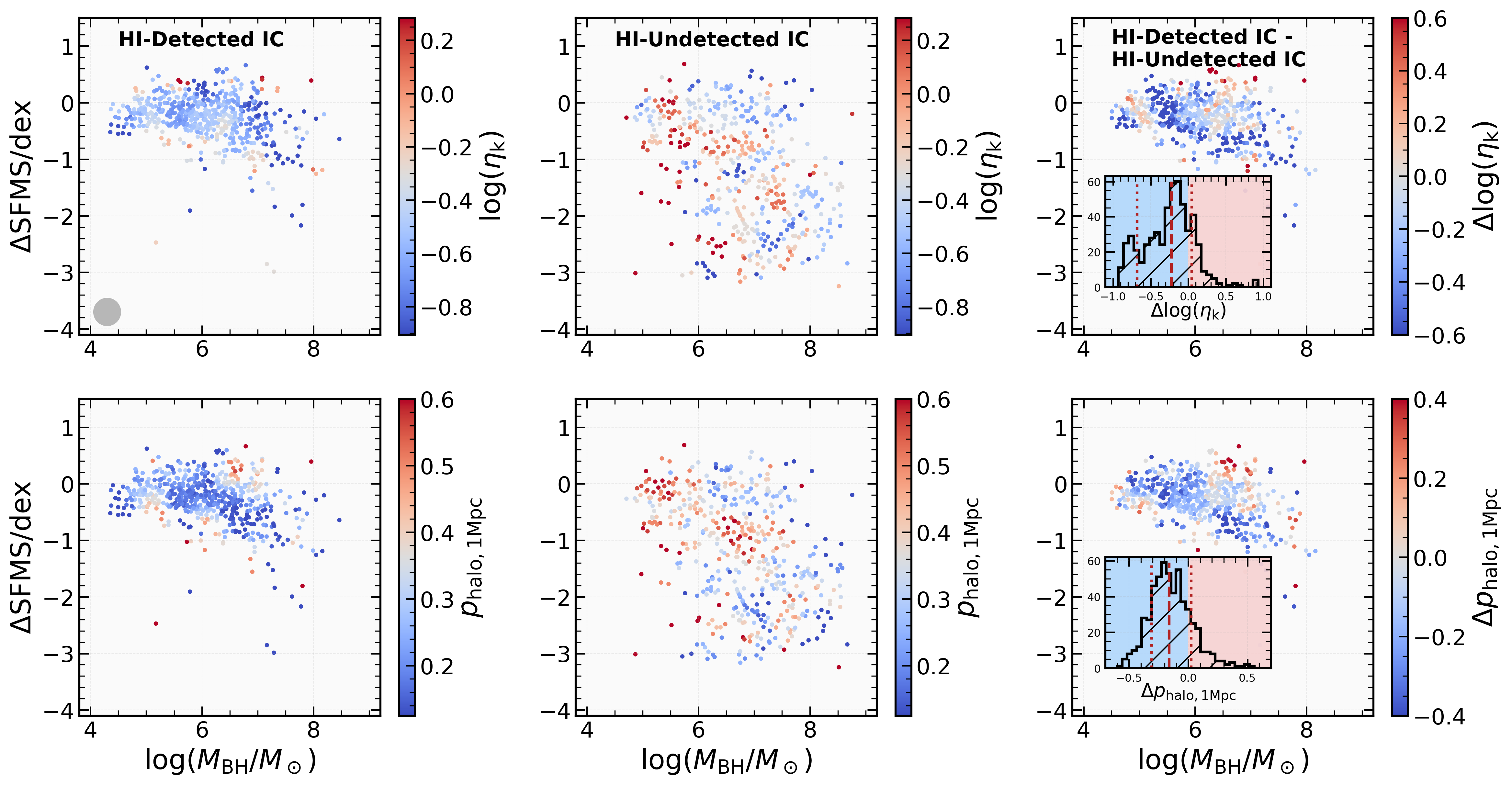}
\caption{\textbf{From the left to the right, we show the large-scale environmental density (upper) and the possibility of finding a nearby group (more massive than $10^{12}M_\odot$) within projected 1\,Mpc (lower) on the $\Delta {\rm SFMS}-M_{\rm BH}$ plane for the \HIt{}-detected ICs, \HIt{}-undetected ICs and their corresponding differences.} The first two columns show the properties as the color coding for the \HIt{}-detected and \HIt{}-undetected ICs. The ellipse in the lower-left of the first panel shows the aperture size used to smooth the color coding. The third column shows the difference between the \HIt{}-detected and \HIt{}-undetected ICs controlling $M_{\rm BH}$ and $\Delta \rm SFMS$. For each \HIt{}-detected galaxy, we identify \HIt{}-undetected counterparts within a 0.25~dex vicinity on the $\Delta\rm SFMS$-$M_{\rm BH}$ plane and calculate the difference to the median value of the matched counterparts. The inserted panel in the third column further shows the 1-D distribution of this color difference derived in the third columns and the red dashed and dotted lines denote the 50th, 16th and 84th percentiles of the distribution.}\label{fig5}
\end{figure*}

\subsection{{The internal reasons for gas deficiency} \red{in ICs}}
After establishing that ICs tend to exhibit \red{lower gas content}{lower median gas fraction} than GCs, here we explore the cause, starting from the internal factors.\\
We first divide ICs into two samples: \HIt{}-detected and \HIt{}-undetected galaxies, \red{and compare their properties}{representing the \HIt{}-normal and \HIt{}-deficit ICs. We note that the sensitivity of \HIt{}-MaNGA survey is relatively shallow and some galaxies that are not as gas poor may be mixed into the \HIt{}-undetected. With deeper and larger \HIt{} surveys, the subsequent results should be more distinguished.}
We then compare their black hole to stellar mass ratio ($M_{\rm BH}/M_\star$), T-Type~\citep[from][]{Dominguez+2018}, and the light-weighted stellar ages within effective radius from MaNGA \textsc{FIREFLY}~\citep{Neumann+2022}. 
{To make a fair comparison, we control the galaxies on the $\Delta {\rm SFMS}-M_{\rm BH}$ plane. This two-dimensional control is essential because both $M_{\rm BH}$ and recent star formation are known to correlate strongly with \HIt{} content. We also perform the control on the $\Delta {\rm SFMS}-M_\star$ plane (which is actually on the star-forming main sequence), the results remain unchanged.}\\
Figure~\ref{fig4} presents the distribution of these properties in \HIt{}-detected galaxies and \HIt{}-undetected galaxies, and the comparative difference between the two samples (and also the one-dimensional distribution of this difference) from the left to the right. The \HIt{}-undetected galaxies exhibit larger $M_{\rm BH}/M_\star$, smaller T-Type, and older stellar population. Early-type galaxies are known to have larger $M_{\rm BH}/M_\star$ than the late-type ones~\citep{Reines+2015,Bosch+2016} as the tight co-evolution is revealed between black holes and bulges~\citep{Kormendy+2013}. Here, although most galaxies on the main sequence ($\Delta{\rm SFMS}\sim 0$) are late-type galaxies with T-Type$ >0$, the \HIt{}-undetected galaxies still have larger $M_{\rm BH}/M_\star$ due to the more bulge-dominant morphology (smaller T-Type) where the supermassive black hole feedback has been shown to be more efficient ~\citep{WangBitao+2023}.  
From an alternative perspective, the black hole  masses and stellar masses serve as indicators of the energy emitted by black holes and the galactic binding energy, respectively~\citep{Bluck+2020,ShiYong+2021}. The larger $M_{\rm BH}/M_\star$ means that the energy released by black holes more easily overcomes the binding energy, which would remove the cool gas within the galaxies. Without further gas accretion from filaments or minor mergers, unlike GCs, these systems are expected to remain gas-poor. We emphasize, however, that the black-hole mass used in this work is an estimator closely tied to the stellar velocity dispersion and structural properties of the galaxies. Morphology-related parameters \citep[e.g., stellar surface density,][]{Catinella+2010,Fabello+2011} could also play a role in the gas deficiency. To assess the residual contribution associated with the black-hole mass proxy, we further perform a multivariate regression analysis controlling for the black-hole mass, T-Type, stellar surface density, velocity dispersion, and $\Delta\rm SFMS$. We find that the residual difference in $\log(M_{\rm BH}/M_\star)$ between \HIt{}-undetected and \HIt{}-detected ICs is 0.09 dex. This relatively small residual highlights the difficulty of disentangling the contribution of black-hole mass from those correlated galaxy properties, given that the black-hole masses are not directly measured in our sample. Consequently, while a black-hole-feedback origin could contribute to the gas deficiency, the participation of correlated structural parameters cannot be fully excluded.\\
Another clear discrimination is that the \HIt{}-undetected ICs have older stellar populations within the effective radius. Given that $\Delta\rm SFMS$ is controlled in Figure~\ref{fig4}, the difference in the ages indicates an early assembly history, rather than a low star-formation level in the \HIt{}-undetected ICs. Those early-formed ICs undergoing long-period evolution without fresh gas supply would appear as gas-deficient galaxies as expected in the strangulation scenario~\citep{PengYingjie+2015}.

\subsection{{The external reasons for gas deficiency} \red{in ICs}}
To evaluate the external effects on the gas deficiency \red{in ICs}, again on the $\Delta{\rm SFMS}-M_{\rm BH}$ plane, we study the environmental density for \HIt{}-detected and -undetected ICs (the upper row in Figure~\ref{fig5}). The environmental density is indicated by the projected density to the 5th nearest neighbor up to 5\,Mpc~\citep[$\eta_{\rm k}=\log\frac{3(k-1)}{4\pi d_{\rm k}^3}$, where $k$ equals 5 and $d_{\rm k}$ is the projected distance to the $k$th nearest galaxy; from][]{Argudo+2015}. \\
Figure~\ref{fig5} shows that the \HIt{}-undetected ICs are located in denser environments than their \HIt{}-detected ones. 
Galaxies in filaments, where the galaxy density is higher than in voids, exhibit a larger quiescent fraction than those in sparser environments~\citep{OKane+2024,Nandi+2025}, and gas deficiency in galaxies becomes more pronounced at smaller distances to the filament spine, indicating a cutoff in the supply of fresh gas~\citep{CroneOdekon+2018}, possibly due to tidal heating. \\
To pinpoint what drives this environmental density excess, for each MaNGA IC we further search in the SDSS group catalog of \cite{Yang+2007} for neighboring groups within $\Delta z<0.01$, and within a projected radius of 1\,Mpc. By incorporating the halo mass ($M_{\rm halo}$) from \cite{Yang+2007}, we calculate the probability of ICs being located in the neighborhood of a galaxy group more massive than $\log(M_{\rm halo}/M_\odot)=12$ on the $\Delta{\rm SFMS}-M_{\rm BH}$ plane (fraction of galaxies with a nearby massive galaxy group at given $M_{\rm BH}$ and $\Delta\rm SFMS$; shown in the lower row of Figure~\ref{fig5}).
\HIt{}-undetected ICs systematically more likely reside in close proximity to a massive halo than \HIt{}-detected ICs at the same $M_{\rm BH}$ and $\Delta \rm SFMS$. This indicates that the higher environmental density of HI-undetected ICs found in Figure~\ref{fig5} is largely attributed to the presence of a nearby massive group or cluster, rather than to a uniform enhancement of the surrounding galaxy number density. In other words, the gas deficiency in these ICs is not merely a consequence of residing in a generic dense region; it is specifically linked to the vicinity of a massive gravitational structure.

A natural question is whether this environmental dependence is specific to ICs or represents a generic property of gas-poor galaxies. First, GCs define the reference $\mu_{\rm HI}-M_{\rm BH}$ relation in \cite{WangTao+2024} and exhibit a higher \HIt{} detection fraction than ICs, indicating that the large-scale environment plays a less significant role. The limited size of the \HIt{}-undetected or gas-deficient GC sample, particularly at $\log(M_{\rm BH}/M_\odot)<6.5$, also restricts similar analyses to that presented in Fig.~\ref{fig5}. Second, even at fixed stellar or black-hole mass, GCs reside in more massive dark-matter halos than ICs and therefore have deeper potential wells, and interactions with satellites also offer as a pathway of gas replenishment absent in ICs. Consequently, GCs are less susceptible to having their cool-gas supply intercepted by a neighboring massive halo. The environmental trends identified here are therefore more pronounced for isolated centrals.

\section{Discussion}\label{sec4}

\begin{figure*}[htbp]
\centering
\includegraphics[width=0.96\textwidth]{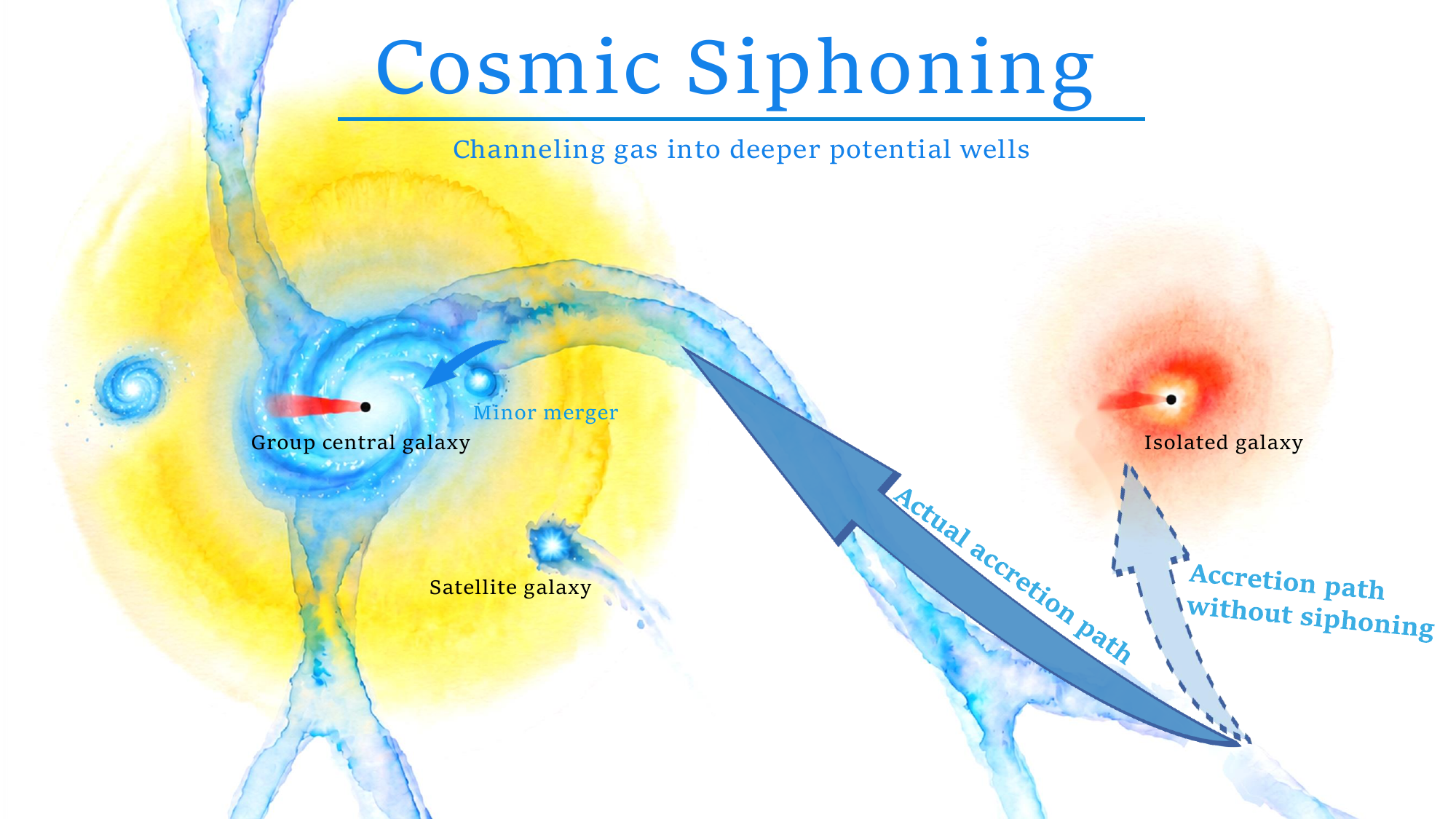}
\caption{
\textbf{
The illustration of ``Cosmic Siphoning" scenario.} The gas flow from the cosmic web toward the isolated central galaxy is interrupted by the massive halo of a nearby galaxy group. Within the group, the minor merger and filaments could continuously replenish gas content in the group central galaxy. And the ram-pressure stripping event would enrich the intra-group medium.}\label{fig_s}
\end{figure*}

The results presented in the previous section suggest that the \red{gas deficiency}{lower median gas fraction} in ICs is not solely an internal phenomenon, and that external mechanisms beyond the host halo \red{may also contribute}{would contribute to the population of gas-deficit ICs}.

Several processes are known to render central galaxies gas-poor. Firstly, galaxies that assembled most of their stars at early epochs tend to be compact, and by the present day have already consumed much of their gas through sustained star formation and outflows, in line with the well established downsizing of galaxy formation~\citep{Cowie+1996,Thomas+2005}. The older stellar ages and earlier morphologies of our \HIt{}-undetected ICs are naturally accommodated by such early assembly without fresh gas supply. Secondly, environments offer further independent routes. Some apparently isolated centrals are actually backsplash systems that have already crossed a group or cluster~\citep{Gill+2005,Mahajan+2011,WangKai+2023,Borrow+2023}, having been stripped and starved during that passage. Even before any group infall, galaxies embedded in cosmic-web filaments may be pre-processed by ram pressure and tides from filaments and from neighboring halos~\citep{Fujita+2004,Kleiner+2017}, which plausibly accounts for the observed \HIt{} deficiency and enhanced quiescence toward filament spines~\citep{Bahe+2013,CroneOdekon+2018,Hoosain+2024}. Finally, part of the offset may reflect halo properties: the rate at which a galaxy accretes gas from the intergalactic medium increases with its halo mass~\citep{Kerevs+2005,Dekel+2006,McBride+2009}, so ICs in lower-mass halos are expected to receive a weaker gas supply even without external interception.

These channels, though diverse in appearance, share a common gravitational origin: the deepest potential wells attract and retain gas at the expense of shallower systems. The preferential proximity of \HIt{}-undetected ICs to massive halos and the higher {median} $\mu_{\rm HI}$ of GCs therefore point to a picture in which massive halos actively regulate the availability of cool gas well beyond their virial radii. We refer to this picture as the ``cosmic siphoning'' effect.

In this scenario, siphoning could operate through two complementary channels. 
On large scales, it behaves as competitive accretion between dark matter halos, where massive halos embedded in the nodes of the cosmic web intercept cold gas flows from surrounding filaments~\citep{Kleiner+2017}. 
Cold-flow accretion along these filaments is recognized as one of the primary modes of gas delivery to halos~\citep{Dekel+2009,Martin+2019}. Lower-mass halos located in the downstream in terms of gas flow are thereby deprived of their fresh fuel and quenched by massive neighbors.  Our results support this picture by directly linking the gas deficiency {in the \HIt{}-undetected IC population} to the presence of a nearby massive halo, which may be the localized driver underlying the \HIt{} deficiency observed toward filament spines earlier.
\\
On smaller, sub-halo scales, an analogous siphoning operates within virialized groups and clusters, where the central galaxy acts as the dominant gravitational sink. Through ram-pressure stripping, tidal interactions, and the accretion of the satellite’s circumgalactic medium~\citep{Merritt+1983,Combes+1988,Moore+1998,Gunn+1972,WangJing+2020,Boselli+2022}, the central galaxy systematically drains gas from its orbiting satellites. This contributes to the elevated \HIt{} content, especially observed in lower-mass GCs~\citep{Janowiecki+2017} before the hot accretion becomes the dominant mode~\citep{Dekel+2006}, while satellites show clear \HIt{} deficiency~\citep{Catinella+2013,Brown+2017,YanShulan+2026}. 

The cosmic siphoning effect does not replace the internal regulation mechanism established in the previous work \citep[e.g., ][]{WangTao+2024}; rather, it acts in concert with it. Once the external gas supply is curtailed, the balance between halo accretion and black hole feedback breaks down. Without fresh gas replenishment, black holes in ICs with higher $M_{\rm BH}/M_\star$ or smaller morphological T-Type could easily heat or expel the remaining cool gas and maintain the gas-deficient state, while the high stellar surface density may also play an effective role~\citep{Kauffmann+2006,Catinella+2010,Fabello+2011}.
On the contrary, GCs sustain a continuously replenished cool gas reservoir through cosmological accretion with more massive halos~\citep{McBride+2009}, tidal-torque-driven inflow from satellite encounters~\citep{Moore+1996,Martinez-Badenes+2012}, and minor mergers. These processes allow for more gradual regulation of gas in GCs under the balance between halos and supermassive black holes. 
Some high-redshift studies also leave clues to this siphoning effect~\citep[e.g.,][]{WangYujan+2025,Ritter+2026}.
We further illustrate this scenario in Figure~\ref{fig_s}.\\
We initiated this work from the comparison between GCs and ICs, thus the GC sample is confined to the $M_{\rm BH}$ (and $M_\star$) distribution of ICs, and GCs in more massive halos are not within the scope of our research. For massive halos, although the accretion rate is higher, the gas accretion is dominated by the hot mode. The accreted gas would be shock-heated and would not directly fuel GCs in the cool-gas phase~\citep[evidence from ][finding the accumulation of cool gas only at larger radii in massive halos]{ZhuRunyu+2026}. However, the cooling efficiency of the hot gas in the center of massive groups/clusters is high, which requires AGN feedback to maintain the non-cool core. This complex balance between gas accretion and AGN feedback would be further discussed in the next paper of this series.\\

\section{Conclusion}\label{sec5}

The aim of this series of papers is to identify the key internal and external processes regulating cool gas accretion in galactic ecosystems. Our previous work demonstrated that supermassive black holes provide the dominant internal regulation mechanism. In this work, we further investigate the secondary role of environmental effects in shaping the cool gas content of central galaxies. Our main findings can be summarized as follows:

\noindent$\bullet$ 
\red{The outliers from the $\mu_{\rm HI}-M_{\rm BH}$ relation are predominantly isolated central galaxies. 
Galaxy locations within the cosmic web (group vs. field) have a significant, independent impact on cold gas accretion and retention, even after accounting for $M_{\rm BH}$.}
{Isolated central galaxies exhibit a higher occupation fraction of outliers from the $\mu_{\rm HI}-M_{\rm BH}$ relation than group centrals, indicating greater diversity in their \HIt{} properties. Group/field environment remains significantly associated with \HIt{} content after accounting for $M_{\rm BH}$, suggesting an additional role for environmental processes in regulating the \HIt{} reservoirs and galaxy evolution of some central galaxies.}

\noindent$\bullet$ 
\HIt{}-undetected isolated galaxies exhibit systematically higher $M_{\rm BH}/M_\star$ ratios and more early-type morphologies. These properties are consistent with a combination of mechanisms that can promote gas depletion, including a reduced ability to retain cool gas and more efficient SMBH feedback, together with the known roles of structural parameters correlated with the black-hole mass proxy. In the absence of fresh gas inflow, such systems are predisposed to remain gas-poor.

\noindent$\bullet$ 
\HIt{}-undetected ICs reside in denser Mpc-scale environments and are preferentially located near massive halos.
This provides \red{direct} observational evidence that external gas supply can be suppressed {in a subset of ICs} by proximity to large-scale structures - likely through stripping and starvation from nearby halo confinement - independent of internal BH feedback.

Taken together, our results support that the regulation of cool gas accretion in central galaxies is governed not only by the internal balance between halos and SMBHs, but also by a cosmic siphoning effect extending from group scales to Mpc scales. Massive halos preferentially attract and retain gas from the surrounding cosmic web, enriching group centrals while depriving nearby isolated galaxies of their gas supply. This external siphoning, when combined with the exhausted effects of AGN feedback, offers a plausible physical interpretation for why the most gas-deficient isolated centrals are preferentially found in the vicinity of massive halos.

\begin{acknowledgments}
{We thank the anonymous referee for his/her insightful comments and constructive suggestions, which helped improve the clarity and interpretation of our results.}
{This work was supported by National Natural Science Foundation of China (grant No.12525302 and 12141301), National Key R\&D Program of China (grant No. 2023YFA1605600), Basic Research Program of Jiangsu (grant No. BK20250001), the Fundamental Research Funds for the Central Universities with grant No.KG202502, and the China Manned Space Program with grant No. CMS-CSST-2025-A04. T.W. acknowledges the support from the New Cornerstone Science Foundation through the XPLORER PRIZE.
}
\red{This work was supported by National Natural Science Foundation of China (Grant No.12525302 and 12141301), Basic Research Program of Jiangsu (Grant No. BK20250001), National Key R\&D Program of China (Grant no. 2023YFA1605600), the Fundamental Research Funds for the Central Universities with Grant no. KG202502, and the China Manned Space Program with grant no. CMS-CSST-2025-A04.}
{We acknowledge the use of Mavis/MiniMax in helping to generate the schematic illustration presented in Figure~\ref{fig_s}.}
\end{acknowledgments}

\begin{contribution}

KW was responsible for writing, data reduction and result presentation.
TW came up with the initial research concept and edited the manuscript.


\end{contribution}

%

\software{astropy~\citep{astropy+2013,astropy+2018,astropy+2022},  
matplotlib~\citep{matplotlib+2007}, 
numpy~\citep{numpy+2020},
reliability~\citep{reliability+2020,reliability+2021},
LtsFit~\citep{Cappellari+2013}
          }

Data availability: the data used in this work is publicly available.


\appendix

\section{xGASS sample and result}\label{appendix}
The galaxy sample selected from xGASS survey follows the same criteria as the MaNGA sample. The group information is from \cite{Janowiecki+2017} based on \cite{Yang+2007}, the stellar mass is from \cite{Salim+2016,Salim+2018}, and the black hole mass is derived from the best-fitted relation of MaNGA sample with 2MASS photometry. Because xGASS galaxies have no direct T-Type information as MaNGA, we cross-match with NASA-Sloan Atlas catalog~\citep{Blanton+2011} for S$\rm\acute{e}$rsic index and categorize ETGs and LTGs based on S$\rm\acute{e}$rsic index with a division at 2.5.

\begin{figure*}[htbp]
\centering
\begin{minipage}[t]{0.99\textwidth}
\includegraphics[width=0.99\textwidth]{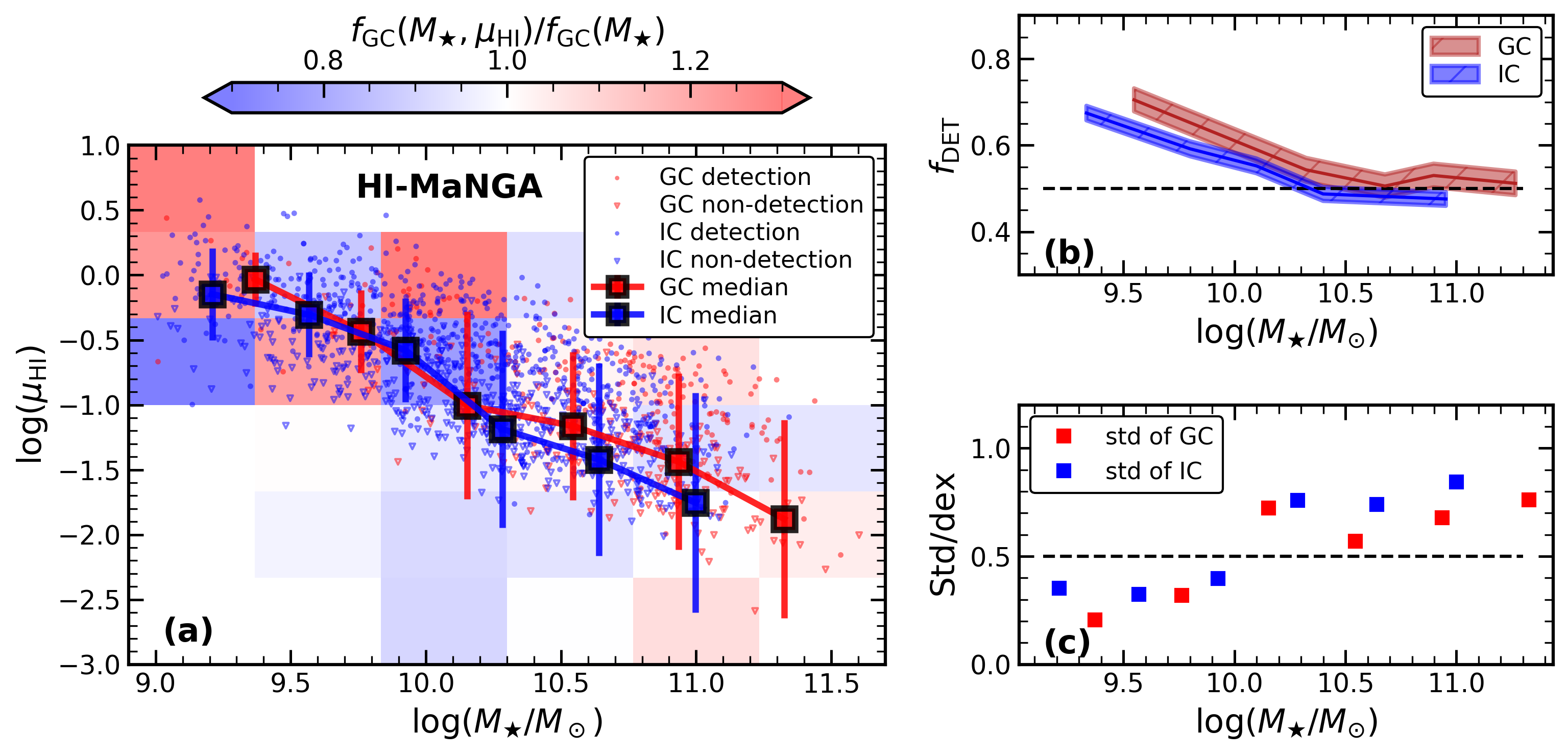}
\end{minipage}
\begin{minipage}[t]{0.99\textwidth}
\includegraphics[width=0.99 \textwidth]{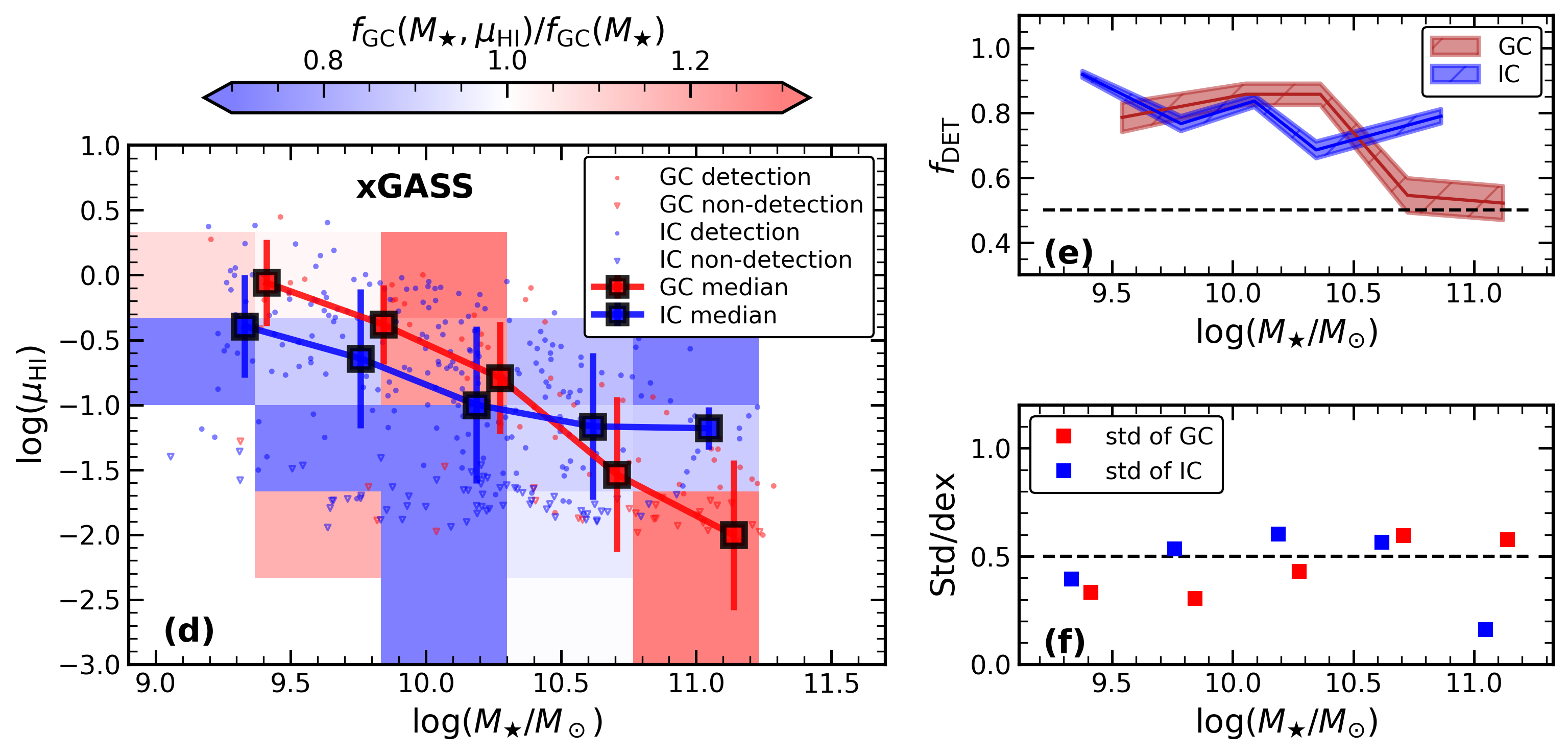}
\end{minipage}
\caption{\textbf{The same as Fig.~\ref{fig2}, but the x-axis is changed to $M_\star$.}}\label{figA1}
\end{figure*}

The $\mu_{\rm HI} - M_{\rm BH}$ and $\mu_{\rm HI} - M_{\star}$ relation with xGASS sample are shown in Fig.~\ref{fig2} and Fig.~\ref{figA1}. 

\begin{figure*}[htbp]
\centering
\begin{minipage}[t]{0.99\textwidth}
\includegraphics[width=0.99\textwidth]{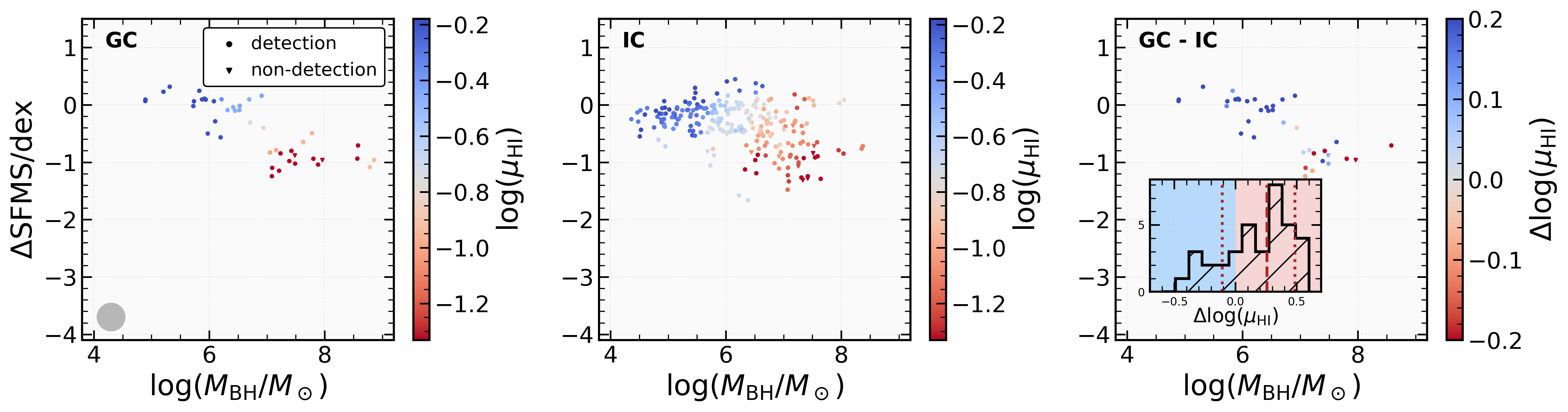}
\end{minipage}
\caption{\textbf{We show the comparison of the gas fraction between ICs and GCs with xGASS sample in the same manner as Fig.~\ref{fig3}.}}\label{figA3}
\end{figure*}

\begin{figure*}[htbp]
\centering
\begin{minipage}[t]{0.99\textwidth}
\includegraphics[width=0.99\textwidth]{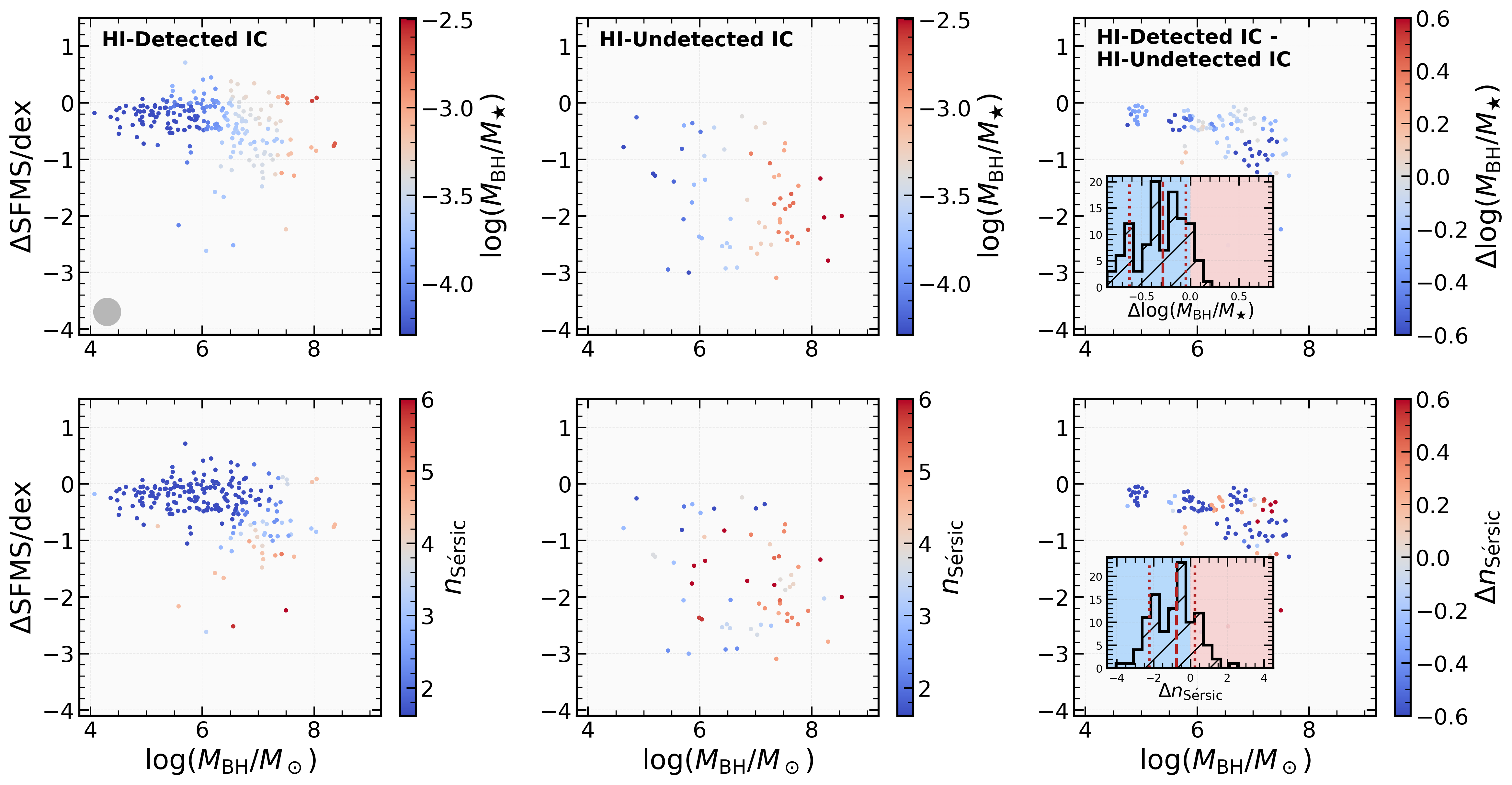}
\end{minipage}
\caption{\textbf{We show the comparison of the $M_{\rm BH}/M_\star$ (upper row) and the S$\rm\acute{e}$rsic index (lower row) between the \HIt{}-detected and undetected galaxies with xGASS sample in the same manner as Fig.~\ref{fig4}.}}\label{figA4}
\end{figure*}

We also compare the $\mu_{\rm HI}$ on the $\Delta \rm SFMS$ and $M_{\rm BH}$ plane in Fig.~\ref{figA3}. Although the sample size is very limited, ICs in xGASS still exhibit lower $\mu_{\rm HI}$ after controlling the $\Delta \rm SFMS$ and $M_{\rm BH}$. When comparing the \HIt{}-detected and undetected ICs in xGASS (Fig.~\ref{figA4}), the \HIt{}-undetected ICs have larger $M_{\rm BH}/M_\star$ ratio and more ETG-like morphology, as we have found in MaNGA sample. The S$\rm \acute{e}$rsic index is larger for \HIt{}-undetected ICs, indicating the more bulge-dominant morphology.

\begin{figure*}[htbp]
\centering
\includegraphics[width=0.99\textwidth]{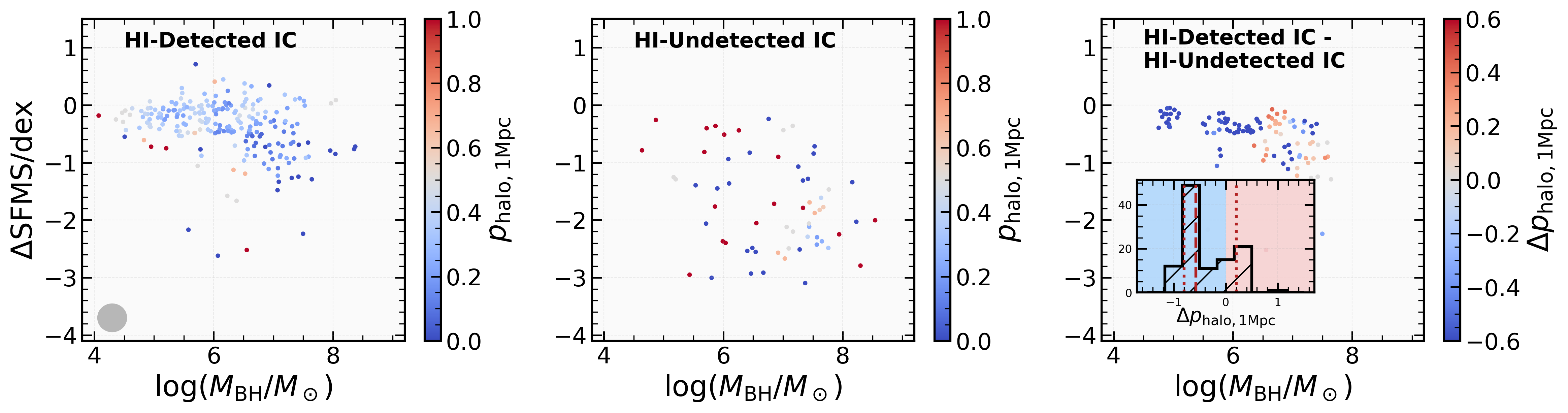}
\caption{\textbf{The same as the lower row in Fig.~\ref{fig5}, but with xGASS sample here.}}\label{figA5}
\end{figure*}

Regarding the large-scale structure around xGASS ICs, we perform the same analysis and also find the excess of probability for \HIt{}-undetected ICs with a nearby massive halo (Fig.~\ref{figA5}).



\bibliography{sample701}{}
\bibliographystyle{aasjournalv7}


\end{CJK*}
\end{document}